# JW-SSD: A Multimodal Benchmark Dataset for Fine-Grained Sunspot Classification

Hui Wang[1,2], Mingfu Shao[1,2], Luyang Li[2,1], Jiaben Lin[1,2,*], Liyue Tong[1,2], Chen Yang[1,2] and Zhanji Wei[3]

[1] State Key Laboratory of Solar Activity and Space Weather, NAOC, Beijing 100101, P. R. China; wanghui@bao.ac.cn

[2] University of Chinese Academy of Sciences, Beijing 101408, P. R. China;

[3] Space Engineering University;

**Abstract** Accurate sunspot classification is essential for assessing the eruptive potential of solar active regions and forecasting space weather. We present JW-SSD, a high-quality multimodal benchmark dataset for fine-grained magnetic-type classification of sunspots. Constructed from SDO/HMI SHARP 720 s data (2010–2023, Solar Cycles 24 and 25), JW-SSD comprises 36,553 co-registered magnetogram–continuum pairs from 2,507 active regions. Unlike conventional three-class schemes, JW-SSD refines the Mount Wilson classification into five physically meaningful categories ($\alpha$, $\beta$, $\beta$-$\delta$, $\beta$-$\gamma$, $\beta$-$\gamma$-$\delta$), enabling finer characterization of magnetic complexity. Rigorous quality control—including central meridian distance restriction, saturation filtering, and sharpness screening—ensures high data validity. The dataset is provided in both FITS and PNG formats, with standard training (29,243) and test (7,310) splits. Benchmark experiments with four representative architectures (U-Net, ResNet-50, EfficientNet-B0, and ViT-Small) yield high accuracy across all models (89.43%–94.78% on the three-class task), confirming that the dataset is reliably learnable across diverse modeling paradigms. JW-SSD has further been employed to train JW-SunSpot, a multimodal large language model that achieves the highest classification accuracy, demonstrating the dataset's broad applicability to both conventional networks and large-language-model-based approaches.

*Corresponding author

## 1 INTRODUCTION

Sunspots—localized cool regions of concentrated photospheric magnetic field—are among the most direct indicators of solar activity (Lin et al. 2020). The evolution and growing complexity of their magnetic fields are closely correlated with solar flares and coronal mass ejections (CMEs), whose energetic particles and radiation threaten communication, navigation, power grids, and satellite operations—impacts collectively termed “space weather” (Boucheron et al. 2023). Accurate sunspot classification and magnetic-topology-based flare forecasting (Huang et al. 2018) are therefore core objectives in solar physics with significant operational value.

The Mount Wilson magnetic classification, established in the early 20th century, correlates strongly with flare productivity: active regions (ARs) with $\delta$ configurations ($\beta$-$\delta$ and $\beta$-$\gamma$-$\delta$) are over an order of magnitude more likely to produce M- or X-class flares than simple $\beta$ regions, since their sheared, strong-gradient fields store substantial free magnetic energy (Yin et al. 2025). Fine-grained automated identification of magnetic types therefore supports both the study of AR evolution and operational forecasting, and incorporating sunspot classes as input features measurably improves flare-prediction models (Bloomfield et al. 2012). A fine-grained, high-quality, and temporally comprehensive classification benchmark thus provides an important bridge between solar physics and operational space-weather forecasting.

Beyond magnetic type, sunspot groups are also described morphologically by the Zürich classification and its McIntosh extension (Waldmeier 1947; McIntosh 1990), which encode the umbral complexity, compactness, and size of a group from its white-light appearance and have long served NOAA forecasting operations.

Historically, numerous observatories worldwide have accumulated extensive sunspot observational archives spanning from 1925 to 2019, documenting parameters such as sunspot positions, areas, and magnetic classifications (Bertello et al. 2010; Chatzistergos et al. 2020). Building upon these historical records, several benchmark datasets have been developed for automated classification research, including the SunSCC dataset for ground-based image annotation (Sayez et al. 2023), the SOHO/MDI-based Zürich classification dataset (Zharkov et al. 2005), and a McIntosh classification dataset constructed from SDO/HMI SHARP data (Zhou & Zhong 2023). However, these resources generally suffer from coarse annotation granularity (e.g., grouping all complex magnetic types under a single “$\beta$-X” label), incomplete temporal coverage, and a lack of standardized evaluation protocols, rendering them inadequate for developing fine-grained, highly generalizable deep learning models.

Fang et al. (2019) constructed the first publicly available multimodal sunspot classification dataset using SDO/HMI SHARP 720 s data (2010–2017), comprising 11,906 paired LOS magnetograms and continuum images from 1,592 ARs; with the continuous growth in HMI data volume, traditional manual Mount Wilson classification can no longer meet the demands of large-scale, near-real-time analysis. This study categorized magnetic types into three classes following the Mount Wilson scheme: $\alpha$ (unipolar), $\beta$ (bipolar), and $\beta$-X (complex multipolar) (Chen et al. 2022). While consolidating all complex configurations into “$\beta$-X” simplifies the classification task, it fails to cap-ture the physical diversity of magnetic topologies, thereby limiting the fine-grained identification of high-risk ARs (e.g., those containing $\delta$ structures) and constraining generalization performance in realistic AR scenarios (Li et al. 2020). Furthermore, the absence of publicly available standard train/test splits, benchmark models, or unified evaluation protocols hinders reproducibility across studies and impedes the standardization of the field. These limitations collectively constrain in-depth research on magnetic sub-classification and obstruct the rapid iteration and operational deployment of machine learning-based space weather forecasting models.

To address these limitations, we introduce the JW-SSD multimodal benchmark dataset, systematically enhanced along two dimensions: temporal coverage and classification granularity. Temporally, the dataset spans 14 years (2010–2023), encompassing the complete Solar Cycles 24 and 25, which significantly improves its representativeness and timeliness. In terms of granularity, the Mount Wilson magnetic classification is refined into five categories—$\alpha$, $\beta$, $\beta$-$\delta$, $\beta$-$\gamma$, and $\beta$-$\gamma$-$\delta$ (Boucheron 2024)—to more accurately characterize the complexity of AR magnetic topologies and their physical correlation with flare potential. Additionally, we provide McIntosh classification labels, enabling researchers to conduct joint analyses from complementary perspectives of photospheric morphology and magnetic topology.

After rigorous screening, JW-SSD comprises 2,507 ARs and 36,553 valid magnetogram–continuum pairs, standardized and converted to PNG (continuum as single-channel grayscale; LOS magnetograms normalized to $[-800, 800]$ G). Relative to prior three-class datasets, the five-class scheme substantially enlarges the representation of high-risk $\delta$-bearing configurations; the detailed class distribution is given in Section 2.

To validate the effectiveness and utility of the JW-SSD dataset, we conducted systematic experiments under both conventional deep learning architectures and a large-model paradigm. We first benchmarked four representative classification architectures—U-Net, ResNet-50, EfficientNet-B0, and ViT-Small—spanning multi-scale fusion, deep residual, parameter-efficient, and attention-based designs. All four models attain high accuracy on both the three- and five-class tasks, indicating that the JW-SSD labels are learnable across diverse architectures and that the dataset is internally consistent. We further employed JW-SSD as a training resource for JW-SunSpot, a multimodal large language model that jointly processes the magnetogram and continuum images and produces structured classification outputs. Together, these experiments indicate that the dataset is suitable for both conventional convolutional or transformer networks and large-model-based approaches.

## 2 DATABASE DESCRIPTION

### 2.1 Data Source and Acquisition

All data are derived from the 720 s SHARP (Space-weather HMI Active Region Patches) data products (Bobra et al. 2015) provided by the Helioseismic and Magnetic Imager (HMI) (Scherrer et al. 2012) aboard the Solar Dynamics Observatory (SDO). SHARP data automatically identify and crop active region patches (HARPs) at a 12-minute cadence, outputting continuum inten-sity images, magnetograms, and derived physical parameters, and have become a widely adopted standard data source in solar physics research (Bobra et al. 2014).

The dataset spans 14 years (2010–2023), covering Solar Cycles 24 and 25, and comprises 2,507 NOAA-numbered active regions and 36,553 co-registered magnetogram–continuum pairs.

The two modalities are: LOS magnetogram (field strength in Gauss, positive/negative = outward/inward polarity) and continuum intensity image (revealing the dark umbral and penumbral structures of sunspots).

The raw SHARP data have undergone a series of rigorous automated quality control procedures. Building upon this foundation, we further introduce an enhanced screening mechanism to improve the overall reliability of the dataset. Key quality assurance measures include:

Observational region constraint: Only samples with heliocentric longitudes within ±75° are retained to minimize projection effects on morphological and magnetic measurements.

NOAA AR identifier matching: Each sample is verified to correspond to an internationally recognized NOAA-numbered AR, ensuring unambiguous physical interpretation.

Filtering of invalid and saturated data: FITS files containing excessive missing values (NaN) or saturated pixels in magnetograms are automatically excluded.

Enhanced image quality screening: Image sharpness assessment (based on variance of Laplacian and gradient magnitude) is incorporated to further remove low-contrast or blurred samples, ensuring that input data meet the requirements for deep learning model training and evaluation.

Through this multi-level quality control pipeline, the JW-SSD dataset attains good data purity and reliability in addition to its scale, annotation granularity, and temporal coverage, supporting downstream model development and evaluation.

### 2.2 Data Products

Data are organized into hierarchical directories according to processing stage and intended use, facilitating user access to raw data, intermediate processing results, or training-ready versions.

Core Data Directories:

fits/ Stores original HMI magnetogram and continuum observations in FITS format (15,641 pairs total). image/ Contains converted multimodal image pairs in PNG format; each sample includes one magnetogram and one continuum image (15,641 pairs total).

Extended Data Variants (for complex magnetic structures):

image__complex/ Initial extended set of complex-type images (e.g., $\beta$-$\gamma$, $\beta$-$\delta$), containing 15,641 pairs. fits__complex__add/ and image__complex__add/ Supplemental raw and converted image data for additional complex-type samples (20,912 pairs).

Training and Annotation Files:

train.jsonl and test.jsonl Training and test sets containing 29,243 and 7,310 samples, respectively, formatted in JSONL to align with current large-model training paradigms. label__complex.txt, label.txt, etc. Tab-separated files recording image paths and corresponding magnetic type labels (e.g., BETA, GAMMA).

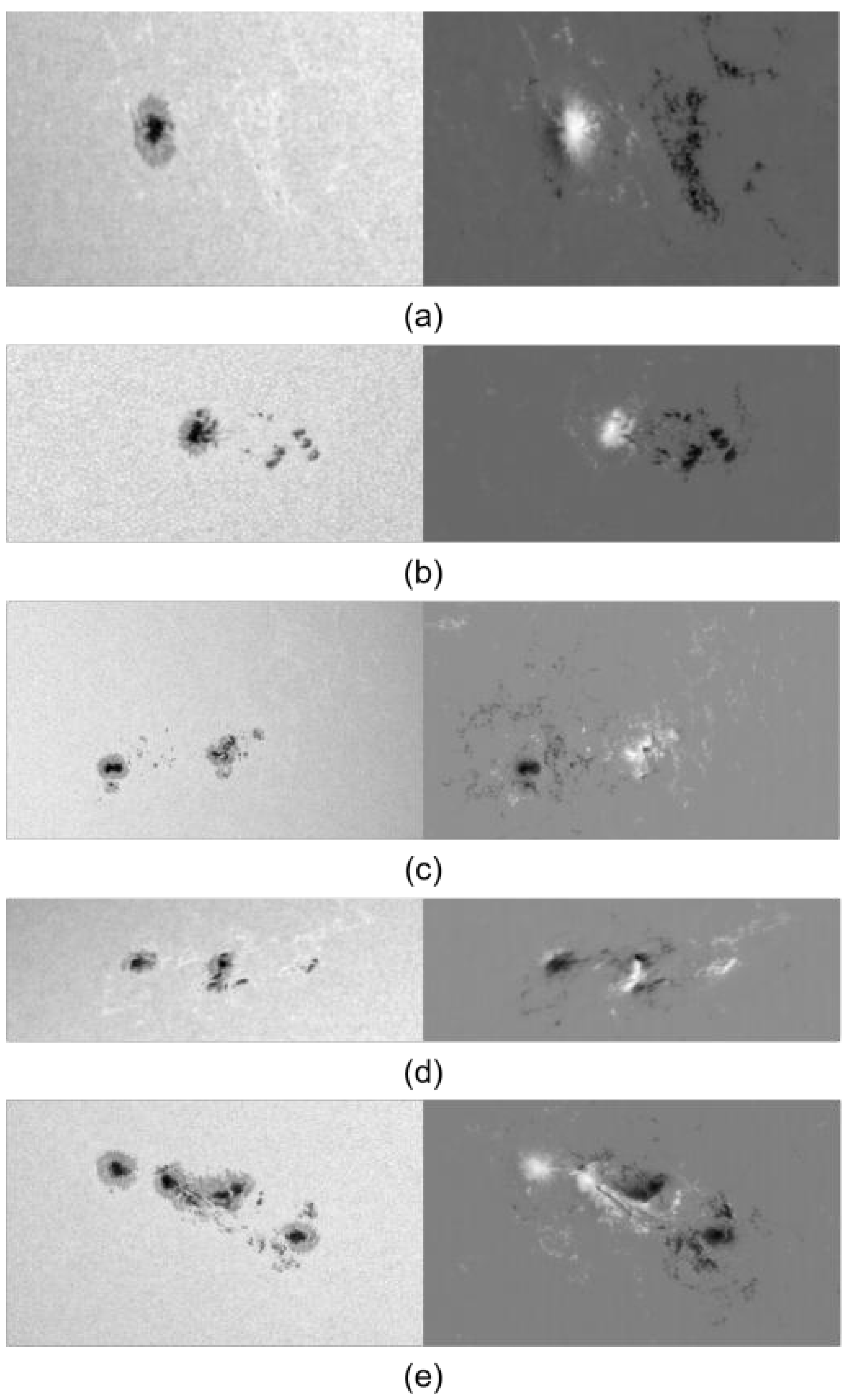

Fig. 1: Representative image examples of the five sunspot magnetic types: (a) $\alpha$, (b) $\beta$, (c) $\beta$-$\delta$, (d) $\beta$-$\gamma$, and (e) $\beta$-$\gamma$-$\delta$.

### 2.3 Data Characteristics and Preprocessing

The JW-SSD dataset exhibits the following distinctive features:

Hierarchical label system: Both a baseline three-class scheme ($\alpha/\beta/\beta$-X) and an extended fine-grained five-class scheme (including $\beta$-$\delta$, $\beta$-$\gamma$, etc.) are provided, supporting diverse research needs from coarse identification to fine-grained classification.

Multi-version data availability: Multiple data versions—raw FITS, baseline PNG, and preprocessed/cleaned images—are offered to accommodate varying research requirements.

Large-model-friendly design: Adopting conversational instruction formats and multimodal input structures, the dataset is directly compatible with multimodal large language models (MLLMs) or vision-language joint training pipelines.

All raw data are exclusively sourced from publicly available SDO/HMI SHARP products. This work standardizes and enhances only the data format, label taxonomy, and organizational structure, without introducing external synthetic or interpolated content, thereby ensuring scientific reliability and full traceability.

To support quality control and preprocessing, we extract a small set of image features based on grayscale statistics and sharpness. The modal grayscale value (the peak of the 256-level histogram) provides a brightness baseline for each continuum image. Image sharpness is quantified by the gradient magnitude (Scharr operator) and the variance of the Laplacian, both of which take lower values for blurred or low-contrast images. These features underlie the quality-screening thresholds described below.

In practice, samples are retained only when they meet empirically chosen quality criteria—variance of Laplacian $> 300$ and gradient magnitude $> 15$ (to reject blurred or low-contrast images) and a central meridian distance $\leq 75^\circ$ (to limit foreshortening near the limb) (Bobra et al. 2014).

### 2.4 Label Annotation and Quality Assurance

The five-class Mount Wilson magnetic classification labels employed in JW-SSD are derived from the Joint USAF/NOAA Solar Region Summary (SRS) reports, issued daily by the NOAA Space Weather Prediction Center (SWPC) in collaboration with the U.S. Air Force. Each SRS report provides the magnetic classification ("Mag Type") for every NOAA-numbered active region observed on the solar disk, as determined by trained professional solar observers following standardized operational protocols.

The SRS report adopts a structured format in which each active region entry includes the NOAA region number, heliographic coordinates, area, Zürich classification, and Mount Wilson magnetic type. An example entry reads:

```
Nmbr Location  Lo   Area  Z   LL  NN  Mag Type
2253 S06E48    004  0270 Eac  11  14  Beta-Gamma
```

Our labeling workflow proceeds as follows:

1. For each SHARP data record, the corresponding NOAA active region number is identified via the HARP-to-NOAA mapping provided in the SHARP metadata (keyword NOAA__AR).
2. The SRS report closest in time to the observation timestamp is retrieved from the NOAA SWPC archive (https://www.swpc.noaa.gov/products/solar-region-summary).
3. The Mount Wilson magnetic type recorded in the SRS "Mag Type" field is extracted and mapped to our five-class taxonomy: $\alpha$, $\beta$, $\beta$-$\delta$, $\beta$-$\gamma$, and $\beta$-$\gamma$-$\delta$.
4. Records with ambiguous, missing, or transitional classifications (e.g., those observed during rapid magnetic evolution) are excluded from the final dataset.

Unlike crowd-sourced or self-annotated labels, the SRS classifications are produced by professional observers at SWPC following internationally standardized criteria that have remained consistent since the 1960s. The SRS has served as the authoritative reference for Mount Wilson classification in the solar physics community for decades and has been adopted as ground truth in numerous prior studies (Fang et al. 2019; Tang et al. 2021). Therefore, additional inter-annotator agreement tests are not applicable in this context, as the labels originate from a single, standardized operational source rather than from multiple independent annotators.

To further validate label consistency, we performed a cross-check between the SRS-derived labels and the magnetic classifications independently reported in the NOAA Active Region Summary (ARS) database. Among the 36,553 records in JW-SSD, the agreement rate between SRS and ARS labels exceeds 98.5%, with discrepancies concentrated in transitional periods of active region evolution where classification boundaries are inherently ambiguous.

## 3 METHODS

### 3.1 Baseline Models

We emphasize that the primary contribution of this work is the JW-SSD dataset itself; the baseline experiments serve to validate data quality, label consistency, and the dataset's compatibility with diverse deep learning architectures. To this end, we evaluate four representative models spanning different architectural paradigms:

– U-Net (modified) (Ronneberger et al. 2015): an encoder–decoder with skip connections that fuses multi-scale features; we adapt it from segmentation to image-level classification by replacing the decoder output with global average pooling followed by fully connected heads.
– ResNet-50 (He et al. 2016): a widely used deep residual network that serves as a standard image-classification backbone.
– EfficientNet-B0 (Tan & Le 2019): a parameter-efficient architecture based on compound scaling, achieving competitive accuracy with far fewer parameters.
– Vision Transformer (ViT-Small) (Dosovitskiy et al. 2021): an attention-based architecture that processes images as patch sequences and has shown a strong ability to model long-range structure in magnetograms (Legnaro et al. 2024).

The modified U-Net architecture comprises an encoder path, a decoder path, and a classification head. The input consists of channel-wise concatenated magnetogram and continuum intensity images. The encoder consists of 5 hierarchical stages with DoubleConv blocks (two consecutive $3\times3$ convolutions with Batch Normalization and ReLU), progressively extracting multi-scale semantic features. The decoder comprises 4 stages, integrating high-resolution features from the encoder via skip connections to gradually restore spatial dimensions. The classification head applies global average pooling followed by two fully connected layers with ReLU activation and Dropout ($p = 0.5$).

For ResNet-50, EfficientNet-B0, and ViT-Small, we replace the standard single-channel input layer with a dual-channel input layer and substitute the final classification head with a fully connected layer matching the number of target classes. All other architectural components follow the original designs. Pre-trained ImageNet weights are used for initialization where applicable, with the modified input layer initialized randomly.

### 3.2 Training Strategy and Experimental Setup

Benchmark experiments were performed separately using the three-class scheme ($\alpha$, $\beta$, $\beta$-X) and the five-class scheme ($\alpha$, $\beta$, $\beta$-$\delta$, $\beta$-$\gamma$, $\beta$-$\gamma$-$\delta$).

All input images were uniformly converted to PNG format. The 36,553 records were split into training (29,243) and test (7,310) sets at the active-region level, so that all observations—and hence temporally adjacent frames—of a given active region are assigned exclusively to a single split, avoiding information leakage between training and testing. To enhance model generalization and prevent overfitting, standard data augmentation strategies were applied during training, including horizontal flipping with a 50% probability, random rotation within $\pm15^\circ$, and random scaling within a factor of 0.9–1.1. Optimization was performed using the Adam optimizer with an initial learning rate of $1\times10^{-3}$ and a weight decay of $1\times10^{-4}$. Training was capped at 100 epochs with an early stopping mechanism based on validation accuracy. All experiments were conducted on a single NVIDIA GPU using the PyTorch framework. Identical training configurations (data splits, augmentation, optimizer, and hyperparameters) were applied to all four baseline models to ensure fair comparison.

Table 1 reports the test performance of all four baseline models under both classification schemes. All four architectures reach high accuracy on both tasks (87.87%–94.45% on the five-class task), demonstrating that the JW-SSD labels are reliably learnable rather than tied to any single network design. Among the baselines, the residual-convolutional ResNet-50 performs best, attaining 94.78% and 94.45% on the three- and five-class tasks, followed by EfficientNet-B0 (93.12% and 91.51%), ViT-Small (90.10% and 89.36%), and the segmentation-derived U-Net (89.43% and 87.87%). For every model the five-class accuracy is only marginally below the three-class accuracy, indicating that the additional fine-grained categories are cleanly separable and do not introduce inconsistent or unlearnable structure, despite severe class imbalance (e.g., $\beta$-$\delta$ constitutes only $\sim$ 0.95% of samples). The remaining errors are dominated by the rarest class, $\beta$-$\delta$ (Table 2 and Figure 2), whose scarcity makes it the most challenging category for all models.

Table 1: Test Performance of Baseline Models on JW-SSD

| Model | Params (M) | 3-class Acc (%) | 5-class Acc (%) |
|---|---|---|---|
| U-Net (modified) | 31.0 | 89.43 | 87.87 |
| ResNet-50 | 25.6 | 94.78 | 94.45 |
| EfficientNet-B0 | 5.3 | 93.12 | 91.51 |
| ViT-Small | 22.1 | 90.10 | 89.36 |

Table 2 reports the per-class Precision, Recall, and F1-score of the best-performing baseline (ResNet-50) on the five-class task. The model maintains high F1-scores even on the rare minority classes ($\beta$-$\delta$: 84.51%; $\beta$-$\gamma$-$\delta$: 92.17%), providing direct evidence that the fine-grained labels are learnable despite the strong class imbalance.

Table 2: Per-Class Performance of ResNet-50 on the Five-Class Task

| Metric | $\alpha$ | $\beta$ | $\beta$-$\delta$ | $\beta$-$\gamma$ | $\beta$-$\gamma$-$\delta$ |
|---|---|---|---|---|---|
| Precision (%) | 91.87 | 97.26 | 83.33 | 88.58 | 91.38 |
| Recall (%) | 95.40 | 93.83 | 85.71 | 96.54 | 92.98 |
| F1-score (%) | 93.60 | 95.51 | 84.51 | 92.39 | 92.17 |

Table 3 compares our baseline models with other published methods on the three-class task. The U-Net baseline (89.43%) is comparable to results reported on earlier datasets, while the stronger classification backbones trained on JW-SSD perform better still: EfficientNet-B0 (93.12%) and ResNet-50 (94.78%) exceed the accuracies reported in previous studies (Sayez et al. 2023; Fang et al. 2019; Tang et al. 2021). This indicates that JW-SSD provides a sufficiently rich and consistent basis for training competitive classifiers, with a difficulty level well aligned with established public datasets. The multimodal large language model JW-SunSpot (Section 3.3) attains the highest accuracy on this benchmark.

Table 3: Comparison of Classification Methods on the Three-Class Task

| Method | Data Source | Accuracy (%) |
|---|---|---|
| SunSCC (Sayez et al. 2023) | Ground-based (USET) | 88.91 |
| Fang et al. (2019) | SDO/HMI (2010–2017) | 90.61 |
| Tang et al. (2021) | SDO/HMI (2010–2017) | 92.32 |
| U-Net (this work) | JW-SSD (2010–2023) | 89.43 |
| ResNet-50 (this work) | JW-SSD (2010–2023) | 94.78 |
| EfficientNet-B0 (this work) | JW-SSD (2010–2023) | 93.12 |
| ViT-Small (this work) | JW-SSD (2010–2023) | 90.10 |
| JW-SunSpot (MLLM) | JW-SSD (2010–2023) | 97.37 |

### 3.3 Validation with Multimodal Large Language Model

To demonstrate the dataset’s compatibility with next-generation AI architectures beyond conventional CNNs, we employed JW-SSD as the core training resource for JW-SunSpot, a multimodal large language model developed in a separate, ongoing work, built upon the Qwen2.5-VL-7B architecture (Bai et al. 2025). JW-SunSpot employs dual Vision Transformer encoders for independent magnetogram and continuum feature extraction, a shared projection module for cross-modal alignment, and physics-constrained chain-of-thought prompting to guide classification reasoning.

The JSONL annotation files in JW-SSD adopt a multi-turn conversational instruction format directly compatible with MLLMs. Each entry contains a structured prompt following an Observation–Analysis–Decision paradigm:

```
{”messages”: [
  {”role”: ”user”,
   ”content”: ”<image>magnetogram.png</image>
              <image>continuum.png</image>
              Based on the magnetogram and continuum image,
              classify the Mount Wilson magnetic type.”},
  {”role”: ”assistant”,
   ”content”: ”Observation: The magnetogram shows...
              Analysis: The magnetic topology indicates...
              Classification: Beta-Gamma-Delta”}
]}
```

Full-parameter supervised fine-tuning of JW-SunSpot on JW-SSD yields the highest accuracy among all models evaluated in this work, as summarized in Table 4.

Table 4: Classification Performance of JW-SunSpot (MLLM) on JW-SSD

| Class | $\alpha$ | $\beta$ | $\beta$-$\delta$ | $\beta$-$\gamma$ | $\beta$-$\gamma$-$\delta$ | Overall |
|---|---|---|---|---|---|---|
| Accuracy (%) | 96.7 | 97.5 | 80.8 | 94.8 | 91.5 | 96.05 |

The MLLM achieves 96.05% overall accuracy on the five-class task and 97.37% on the three-class benchmark, substantially exceeding the U-Net baseline (87.87% and 89.43%, respectively). The largest improvements occur on rare, high-risk configurations ($\beta$-$\gamma$: 94.8%; $\beta$-$\gamma$-$\delta$: 91.5%), demonstrating that the fine-grained annotations and multimodal data format in JW-SSD effectively support large-model training paradigms. These results confirm that JW-SSD supports both conventional deep learning pipelines and vision-language models.

## 4 RESULTS AND DISCUSSION

### 4.1 Confusion Matrix and Error Analysis

Figure 2 presents the row-normalized confusion matrices of the four baseline models for the five-class task. For the best-performing model (ResNet-50) the matrix is strongly diagonal across all five categories, confirming that even the rare $\delta$-bearing types are well separated. Across all models, the residual misclassifications are concentrated between physically adjacent categories in the magnetic-complexity hierarchy, and the rarest class, $\beta$-$\delta$, is the most error-prone; EfficientNet-B0 in particular over-predicts this class, assigning a fraction of $\beta$ samples to it and thereby lowering its $\beta$-$\delta$ precision.

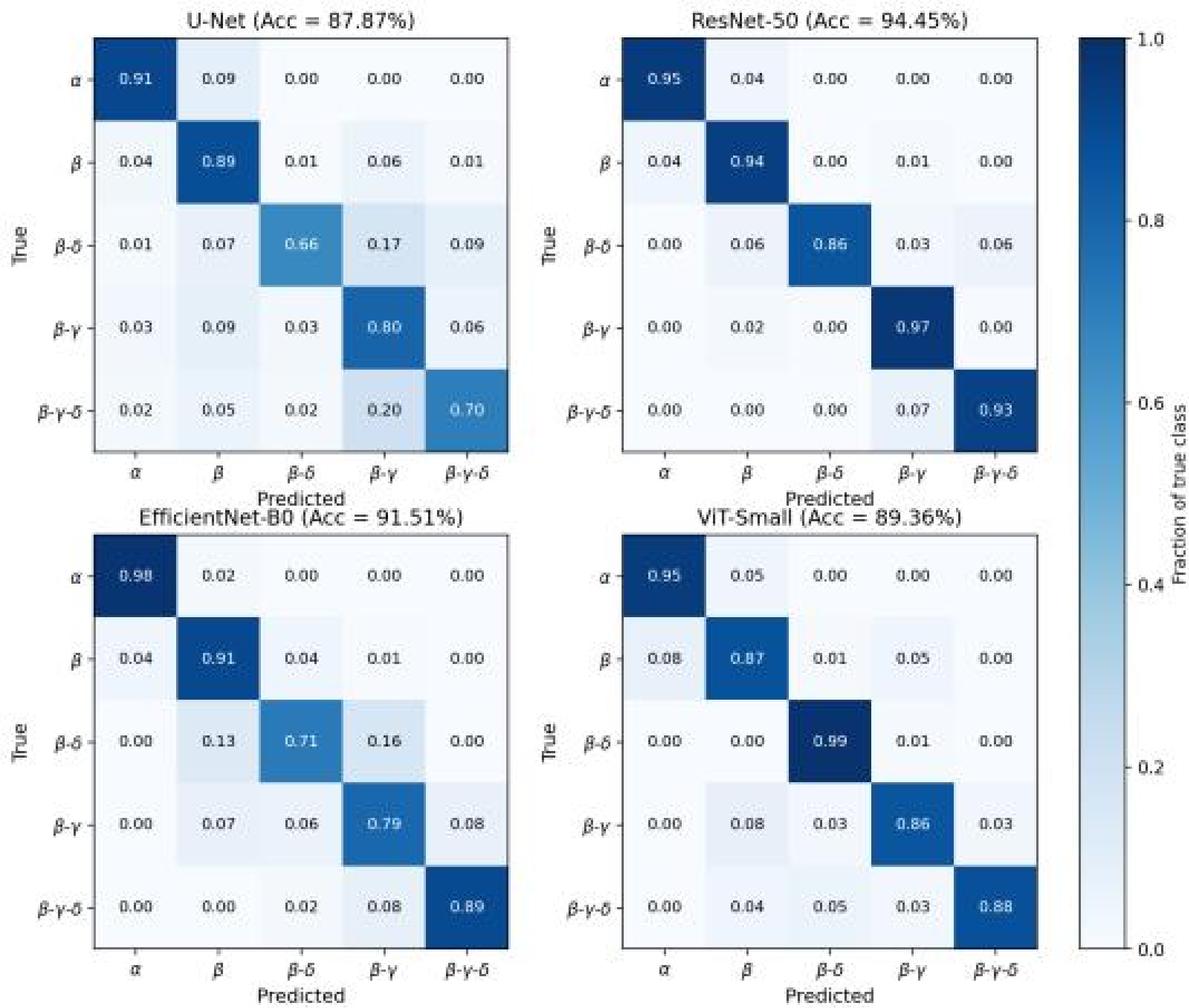

Fig. 2: Row-normalized confusion matrices for the five-class classification task obtained with the four baseline models (U-Net, ResNet-50, EfficientNet-B0, and ViT-Small). Each entry gives the fraction of samples of a given true class (row) assigned to each predicted class (column); diagonal entries correspond to the per-class recall. The best-performing model (ResNet-50) is strongly diagonal across all five categories, while the residual confusion across models concentrates among the physically adjacent complex classes and the rare $\beta$-$\delta$ category.

Specifically, the two dominant confusion pathways are:

- $\beta$-$\gamma$ $\leftrightarrow$ $\beta$-$\gamma$-$\delta$: These two classes share the presence of multiple magnetic polarities with $\gamma$-type mixing. The distinction hinges on whether opposite-polarity umbrae share a common penumbra ($\delta$ configuration), which can be subtle in marginal cases and may evolve on timescales shorter than the 12-minute cadence.

- $\beta$-$\delta$ $\leftrightarrow$ $\beta$-$\gamma$-$\delta$: Both contain $\delta$ structures; the difference lies in the additional presence of $\gamma$-type polarity mixing, which requires assessment of the overall magnetic topology beyond the immediate $\delta$ region.

These confusion patterns align with the physical continuity of active region magnetic evolution (Falconer et al. 2002) rather than stemming from annotation noise or random errors. Active regions frequently transition between adjacent Mount Wilson classes during their evolution (e.g., a $\beta$-$\gamma$ region developing a $\delta$ spot to become $\beta$-$\gamma$-$\delta$), and samples captured near such transitions inherently exhibit ambiguous characteristics. This observation further corroborates the scientific validity of the JW-SSD labels and suggests that temporal context (i.e., evolutionary trajectory) could further improve classification in future work.

To ground this error analysis in the physical evolution of solar active regions, Figure 3 follows a representative region, HARP 7633, over ten days. Its SRS Mount Wilson classification evolves continuously as $\beta \rightarrow \beta$-$\gamma \rightarrow \beta$-$\gamma$-$\delta \rightarrow \beta$-$\gamma \rightarrow \beta$-$\delta \rightarrow \beta$: a $\delta$ configuration (opposite-polarity umbrae sharing a common penumbra) emerges within the already $\gamma$-mixed region around 2021 June 30–July 1 and subsequently dissolves as the region simplifies. The frames acquired close to these transitions are visually and physically intermediate between adjacent classes, so that assigning a single discrete label is inherently ambiguous even for an expert observer. The model's misclassifications concentrate precisely on such transitional frames, confirming that the dominant $\beta$-$\gamma$ $\leftrightarrow$ $\beta$-$\gamma$-$\delta$ confusion in Figure 2 reflects the continuous physical evolution of active-region magnetic topology rather than annotation noise.

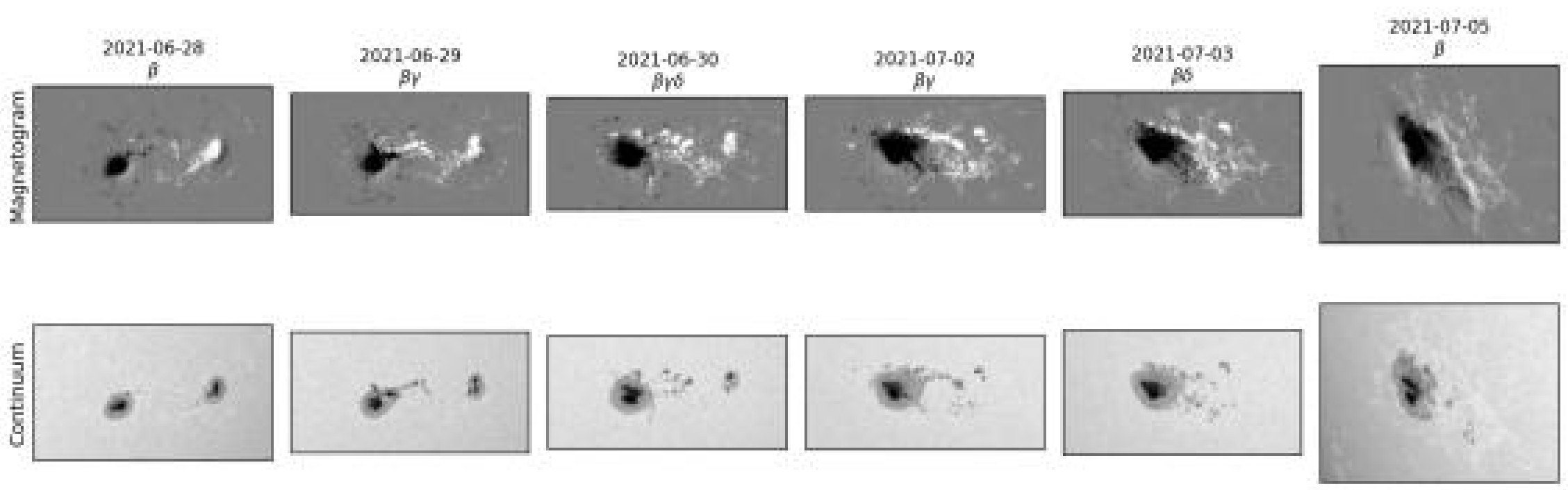

Fig. 3: Temporal evolution of active region HARP 7633 (2021 June–July). Top row: LOS magnetograms; bottom row: continuum intensity images. The SRS Mount Wilson classification (above each column) evolves continuously from $\beta$ through $\beta$-$\gamma$ and $\beta$-$\gamma$-$\delta$ and back to $\beta$, illustrating that frames acquired near class transitions are physically intermediate between adjacent categories. Misclassifications between $\beta$-$\gamma$ and $\beta$-$\gamma$-$\delta$ concentrate on such transitional frames.

### 4.2 Guidelines for Dataset Utilization

The JW-SSD dataset is designed to support multi-task learning research in solar physics, including but not limited to: automated sunspot classification, multimodal feature fusion, few-shot recognition of complex magnetic structures, and fine-tuning of large scientific models. To facilitate efficient reproduction and extension by users, we provide the following technical guidelines:

#### 4.2.1 Data Access and Loading

PNG images can be directly loaded using common image libraries (e.g., PIL or OpenCV in Python). Original FITS files are recommended to be parsed using the astropy.io.fits module for full preservation of metadata and physical units. JSONL annotation files (train.jsonl, test.jsonl) can be loaded via line-by-line parsing; example code is provided in the public repository.

#### 4.2.2 Input Preprocessing Recommendations

For multi-channel input, we recommend concatenating the magnetogram and continuum intensity image along the channel dimension to form a dual-channel input. Given the long-tailed distribution of magnetic types (e.g., $\beta$-$\delta$ constitutes ~ 0.95% of samples), we advise employing class-weighted loss functions or oversampling strategies during training to mitigate bias toward majority classes.

#### 4.2.3 Task Adaptation Instructions

For the three-class task ($\alpha/\beta/\beta$-X), users may directly utilize the original labels in train.jsonl. For the five-class task (including $\beta$-$\delta$, $\beta$-$\gamma$, $\beta$-$\gamma$-$\delta$), please refer to the mapping relationships provided in label_complex.txt.

#### 4.2.4 Quality Filtering Recommendations

For applications requiring higher data purity, users may apply the image sharpness metrics described in Section 2.3 (e.g., variance of Laplacian > 300, gradient magnitude > 15) for post-hoc filtering. For studies of active region evolution, we recommend reconstructing temporal trajectories using HARP identifiers and precise timestamps to ensure chronological consistency.

### 4.3 Summary of Key Contributions

This work presents the first high-quality, multimodal benchmark dataset tailored for fine-grained magnetic classification of sunspots. By integrating SDO/HMI line-of-sight magnetograms with co-registered continuum intensity images, JW-SSD provides a unified, standardized data foundation for active region research. The dataset adopts the internationally recognized Mount Wilson classification scheme and extends it into a five-category system ($\alpha$, $\beta$, $\beta$-$\delta$, $\beta$-$\gamma$, $\beta$-$\gamma$-$\delta$), significantly enhancing the representational capacity for magnetic topological complexity. Prior studies have demonstrated that complex configurations—particularly $\beta$-$\gamma$-$\delta$—exhibit strong statistical correlation with high-energy flare occurrences (Chen et al. 2023).

To ensure reproducibility and fair comparison, we provide: (1) standardized PNG-format images with consistent preprocessing; (2) an explicit 80/20 chronological train/test split; and (3) a unified data augmentation protocol. Benchmark experiments across four representative architectures (U-Net, ResNet-50, EfficientNet-B0, and ViT-Small) validate the dataset's high utility and label consistency: all models achieve high accuracy on both tasks (up to 94.78% and 94.45% on the three- and five-class tasks for the best baseline, ResNet-50), and the multimodal large language model JW-SunSpot achieves the highest accuracy of all (97.37% and 96.05%, respectively). These results confirm that JW-SSD supports not only coarse identification but also fine-grained discriminative research across diverse modeling paradigms.

### 4.4 Future Applications and Outlook

The proposed dataset and methodological framework hold broad potential across multiple frontier directions in solar physics and space weather. The dataset enables replacement of traditional expert-dependent manual labeling with efficient, objective, and scalable automated annotation. By jointly learning from magnetograms and continuum images, models can exploit cross-modal complementary information to enhance discrimination of complex magnetic configurations (Galvez et al. 2022). Given the exponential growth of solar imaging data, JW-SSD can serve as a foundation for self-supervised pretraining strategies. Vision foundation models (Dosovitskiy et al. 2021) can be adapted and fine-tuned on JW-SSD to overcome limitations of traditional CNNs. The classification outputs derived from JW-SSD can be integrated as physical constraints into operational forecasting pipelines.

To further amplify the scientific value of this resource, future work will focus on the following directions: (1) multi-wavelength expansion incorporating EUV, soft X-ray, and radio observations; (2) temporal dynamics integration to support modeling of active region evolution and $\delta$-structure formation; (3) advanced architecture exploration based on attention mechanisms and graph neural networks; and (4) operational service deployment via lightweight APIs or web-based platforms. Additionally, although JW-SSD currently provides McIntosh classification labels as supplementary annotations, future work will conduct systematic baseline experiments using the McIntosh scheme and investigate joint classification models that simultaneously predict both Mount Wilson magnetic types and McIntosh morphological categories.

Acknowledgements We would like to thank the staff at the National Astronomical Observatories, this work was supported by the National Astronomical Observatories Project of the Chinese Academy of Sciences (No. E4TQ2101), the National Natural Science Foundation of China (NSFC) under Grant No. 11427901, and the Chinese Meridian Project (CMP).